\documentstyle[twocolumn,aps]{revtex}

\begin{document}
\draft   

\title{General structure of the graviton self-energy} 

\bigskip

\author{F. T. Brandt and J. Frenkel}
\address{Instituto de F\'\i sica,
Universidade de S\~ao Paulo\\
S\~ao Paulo, SP 05315-970, BRAZIL}

\date{\today}
\maketitle 

\medskip

\begin{abstract}
The graviton self-energy at finite temperature depends on fourteen
structure functions. We show that, in the absence of tadpoles, the gauge
invariance of the effective action imposes three non-linear relations
among these functions. The consequences of such constraints,
which must be satisfied by the thermal graviton self-energy to all
orders, are explicitly verified in general linear gauges to one loop
order. 
\end{abstract}
\bigskip
\bigskip

The non-linear relation imposed by gauge invariance on the thermal 
self-energy of gluons, has been recently discussed by Weldon in an
interesting paper \cite{weldon:1996kb}. He proved that in QCD, the
Slavnov-Taylor identities \cite{slavnov:1972fg,taylor:1971ff}
require a non-linear constraint among the structure functions which 
occur at finite temperature. In this brief report, we show 
that a similar behavior occurs in the gauge theory of gravity.
In this case, local gauge invariance leads to three non-linear
relations which restrict the form of the  thermal self-energy 
of gravitons.

The Einstein theory of gravity is described by the Lagrangian density
\cite{gross:1982cv,kikuchi:1984np,gribosky:1989yk}
\begin{equation}
  \label{eq:1}
  {\cal L}=\frac{2}{\kappa^2} \sqrt{-g} R ,
\end{equation}
where $\kappa^2=32\pi G$, $G$ is the Newton constant and $R$ is the
Ricci scalar.
The graviton field $h_{\mu\nu}$ can be defined in terms of
the metric tensor as
\begin{equation}
  \label{eq:2}
  g_{\mu\nu}=\eta_{\mu\nu}+\kappa h_{\mu\nu}.
\end{equation}
Using this parametrization, the Einstein action will be 
invariant under the 
gauge transformation \cite{matsuki:1979rt}
\begin{eqnarray}
  \label{eq:3}
  \delta h_{\mu\nu}& = &
\left[ \delta^\lambda_\mu\partial_\nu+
                           \delta^\lambda_\nu\partial_\mu+
\kappa\left(h^\lambda_\mu\partial_\nu+
            h^\lambda_\nu\partial_\mu+
            \partial^\lambda h_{\mu\nu}\right)\right]\xi_\lambda
\nonumber \\
& \equiv & G^{(0)\,\lambda}_{\mu\nu} \xi_\lambda,
\end{eqnarray}
where $\xi_\lambda$ is an infinitesimal gauge parameter.

In this gauge theory, the corresponding identities which occur at
finite temperature, differ from those at $T=0$ because of the
appearance of one-particle graviton functions (tadpoles). Their
thermal contribution is purely leading, being
proportional to $T^{2(n+1)}$ at the $n$-loop order. Hence, we may
assume that the tadpoles are important, in the Ward
identities, only for the leading thermal contributions to the
graviton self-energy. (To one-loop order, for example,
the tadpoles can be neglected for the purpose
of studying the sub-leading $T^2$, $\log(T)$ and $T=0$ contributions).

At finite temperature, the graviton self-energy may depend on the
four-velocity $u_\alpha$ of the plasma, so that it can be a linear
combination of the 14 independent tensors 
given in Table \ref{tab1}. This 
contains three traceless tensors
$T^A_{\alpha\beta,\, \mu\nu}$, $T^B_{\alpha\beta,\, \mu\nu}$ 
and $T^C_{\alpha\beta,\, \mu\nu}$,
which are transverse with respect to the wave 4-vector $k_\mu$ . 
They are also,
respectively, completely transverse, partially transverse and
longitudinal with respect to the spatial component $\vec k$
\cite{rebhan:1991yr}. These tensors depend individually on the 
plasma four-velocity, but their sum is a Lorentz covariant tensor
which is independent of $u_\alpha$
\begin{eqnarray}
  \label{eq:9}
\left(T^A + T^B + T^C\right)_{\alpha\beta,\, \mu\nu} =
\nonumber \\ 
\frac 1 2 \left(P_{\alpha\mu}P_{\beta\nu}+
                P_{\alpha\nu}P_{\beta\mu}\right) -
\frac 1 3 P_{\alpha\beta}P_{\mu\nu},
\end{eqnarray}
where $P_{\alpha\beta} = \eta_{\alpha\beta} - 
                          k_\alpha k_\beta/k^2$.

In terms of this basis, which is convenient for our purpose,
the graviton self-energy can be parametrized as
\begin{eqnarray}
  \label{eq:4}
\Pi_{\alpha\beta,\, \mu\nu}&=&\Pi_A T^A_{\alpha\beta,\, \mu\nu}+
                            \Pi_B T^B_{\alpha\beta,\, \mu\nu}+
                            \Pi_C T^C_{\alpha\beta,\, \mu\nu}
\nonumber \\
&+&\sum_{i=4}^{14}\Pi_i T^i_{\alpha\beta,\, \mu\nu}
\end{eqnarray}

\begin{table}[h!]
\begin{center}
$$
  \begin{array}{l}
T_{\alpha\beta ,\, \mu\nu}^{1,2,3}
=T^{A,B,C}_{\alpha\beta ,\, \mu\nu}
\\
T_{\alpha\beta ,\, \mu\nu}^4=\eta _{\alpha \beta}\,\eta _{\mu \nu}   \\ 
T_{\alpha\beta ,\, \mu\nu}^5=u_{\mu} \,u_{\nu} \,\eta _{\alpha \beta}+
u_{\alpha} \,u_{\beta} \,\eta _{\mu \nu}  \\ 
T_{\alpha\beta ,\, \mu\nu}^6=\left[u_{\beta} \,\left( k_{\nu} \,
\eta _{\alpha \mu}+k_{\mu} \,\eta _{\alpha \nu}\right) +
k_{\beta} \,\left( u_{\nu} \,\eta _{\alpha \mu}+u_{\mu} \,\eta _{\alpha \nu}
\right) +\right.  \\
\;\;\;\;\;\;\;\;\;\;\;\;\;\;
\left. u_{\alpha} \,\left( k_{\nu} \,\eta _{\beta \mu}+
k_{\mu} \,\eta _{\beta \nu}\right) +k_{\alpha} \,\left( u_{\nu}
\,\eta _{\beta \mu}+u_{\mu} \,\eta _{\beta \nu}\right)\right]/k\cdot u   \\ 
T_{\alpha\beta ,\, \mu\nu}^7=\left[
k_{\nu} \,u_{\alpha} \,u_{\beta} \,u_{\mu} +
k_{\mu} \,u_{\alpha} \,u_{\beta} \,u_{\nu} +\right. \\
\;\;\;\;\;\;\;\;\;\;\;\;\;\;\;
\left.
k_{\beta} \,u_{\alpha}\,u_{\mu} \,u_{\nu} +
k_{\alpha} \,u_{\beta} \,u_{\mu} \,u_{\nu}\right]/k\cdot u
\\ 
T_{\alpha\beta ,\, \mu\nu}^8=
\left[
k_{\beta} \,k_{\nu} \,\eta _{\alpha \mu}+
k_{\beta} \,k_{\mu} \,\eta _{\alpha \nu}+
k_{\alpha} \,k_{\nu} \,\eta _{\beta \mu}+
k_{\alpha} \,k_{\mu} \,\eta _{\beta \nu}\right]/k^2  \\ 
T_{\alpha\beta ,\, \mu\nu}^9=
\left[
k_{\mu} \,k_{\nu} \,u_{\alpha} \,u_{\beta} +
k_{\alpha} \,k_{\beta} \,u_{\mu} \,u_{\nu}\right]/k^2  \\ 
T_{\alpha\beta ,\, \mu\nu}^{10}=\left( k_{\beta} \,u_{\alpha} +
k_{\alpha} \,u_{\beta} \right) \,\left( k_{\nu} \,u_{\mu} +k_{\mu} \,u_{\nu} \right) /(k\cdot u)^2  \\ 
T_{\alpha\beta ,\, \mu\nu}^{11}=
\left[
k_{\beta} \,k_{\mu} \,k_{\nu} \,u_{\alpha} +
k_{\alpha} \,k_{\mu} \,k_{\nu} \,u_{\beta} +\right. \\
\;\;\;\;\;\;\;\;\;\;\;\;\;\;\;
\left.
k_{\alpha} \,k_{\beta} \,k_{\nu} \,u_{\mu} + 
k_{\alpha} \,k_{\beta} \,k_{\mu} \,u_{\nu}\right]/(k^2 \,k\cdot u)\\ 
T_{\alpha\beta ,\, \mu\nu}^{12}=
\left(k_{\alpha} \,k_{\beta} \,k_{\mu} \,k_{\nu}\right)/k^4\\ 
T_{\alpha\beta ,\, \mu\nu}^{13}=
\left(
k_{\mu} \,k_{\nu} \,\eta _{\alpha \beta}+
k_{\alpha} \,k_{\beta} \,\eta _{\mu \nu}
\right)/k^2  \\ 
T_{\alpha\beta ,\, \mu\nu}^{14}=
\left[
\left( k_{\nu} \,u_{\mu} +k_{\mu} \,u_{\nu}
\right) \,\eta_{\alpha \beta}+\left( k_{\beta} \,u_{\alpha} +
k_{\alpha} \,u_{\beta} \right) \,\eta_{\mu \nu}
\right]/(k\cdot u)
 \\
   \end{array}
$$
\smallskip
\caption{A basis of 14 independent tensors.\label{tab1}}
\end{center}
\end{table}

In the hard thermal loop approximation, which represents a consistent
high-temperature expansion
\cite{braaten:1990mz,frenkel:1990br},
the one loop graviton self-energy has leading contributions
proportional to $T^4$, and the corresponding functions 
$\Pi_j^{(l)}$ are gauge invariant \cite{rebhan:1991yr}. However, 
when one goes beyond this approximation, by including contributions
which are sub-leading in powers of $T$, this feature no longer
occurs and the functions $\Pi_j^{({ s})}$ become gauge dependent
\cite{brandt:1998hd}.

In order to investigate the structure of the exact graviton
self-energy which includes also higher loops effects, we will
make use of the Becchi-Rouet-Stora identities \cite{becchi:1974md}
which reflect the underlying gauge invariance of the theory
\cite{capper:1982ez,delbourgo:1985wz}. A discussion of the 
consequences of these identities on the structure of the thermal
self-energy is given in Appendix A. To explain these, we will
denote by $\check\Gamma_{\alpha\beta,\, \mu\nu}$
the quadratic part of the graviton effective action, 
which is the sum of the free kinetic energy, without the gauge
fixing term, and the one-particle-irreducible 
graviton self-energy

\begin{equation}
  \label{eq:5}
\check\Gamma_{\alpha\beta,\, \mu\nu}=K^{(0)}_{\alpha\beta,\, \mu\nu}+
\Pi_{\alpha\beta,\, \mu\nu}.
\end{equation}
The free-graviton kinetic energy is given, in momentum
space, by
\begin{eqnarray}
  \label{eq:6}
& K^{(0)}_{\alpha\beta,\, \mu\nu}(k) =  k^2\left(
                             \eta_{\alpha\mu}\eta_{\beta\nu}+  
                             \eta_{\alpha\nu}\eta_{\beta\mu}-
                             \eta_{\alpha\beta}\eta_{\mu\nu}
                                 \right) 
\nonumber \\
& + 
k_\mu k_\nu \eta_{\alpha\beta} + k_\alpha k_\beta \eta_{\mu\nu} 
\nonumber \\                                
& - 
\left(k_\alpha k_\mu \eta_{\beta\nu} +
      k_\alpha k_\nu \eta_{\beta\mu} +
      k_\beta k_\mu \eta_{\alpha\nu} +
      k_\beta k_\nu \eta_{\alpha\mu} \right)  
\end{eqnarray}

Of course, the contributions to $\Pi_{\alpha\beta,\, \mu\nu}$ are
calculated according to the usual Feynman rules, with a gauge fixing
term in the bare propagator. Hence, the gauge dependence of
$\check\Gamma_{\alpha\beta,\, \mu\nu}$
comes only from the self-energy functions. In consequence of
the BRS identities, it turns out that the leading contributions
to the longitudinal part of $\check\Gamma_{\alpha\beta,\, \mu\nu}$ 
are proportional to the tadpole terms (see Eq. (\ref{eq:A5})).
Furthermore, the
sub-leading contributions to $\check\Gamma_{\alpha\beta,\, \mu\nu}$ 
satisfy the following four constraints (see Eq. (\ref{eq:A6}))
\begin{equation}
  \label{eq:7}
\check\Gamma_{\;\alpha\beta}^{({ s})\,\mu\nu}(k)  
\left.G^\lambda_{\mu\nu}(k)\right|_{h=0} = 0 \;\;\;\;\;\;\;\; 
{\rm for}\;\; \lambda=0,1,2,3 ,
\end{equation}
where the tensor $G^\lambda_{\mu\nu}$ is given,
to lowest order,  by (\ref{eq:3}).

Since $\check\Gamma_{\;\alpha\beta}^{({ s})\,\mu\nu}$ is
symmetric under permutations of indices $\alpha\leftrightarrow\beta$
and $\mu\leftrightarrow\nu$, it can viewed as a $10\times 10$ matrix
which must have $10$ eigenvalues. Six of these are determined
dynamically by the equations of motion of the gravitational
field. The other four eigenvalues must be zero in consequence
of the gauge invariance as expressed by Eq. (\ref{eq:7}).
In order to obtain zero eigenvalues, it is necessary that the
determinant of $\check\Gamma_{\;\alpha\beta}^{({ s})\,\mu\nu}$
should vanish. These requirements lead to three distinct non-linear
constraints among the exact functions 
$\Pi^{({ s})}_j$, which determine the structure of the
graviton self-energy at finite temperature.
The first of these non-linear relations,
which are derived in Appendix B, can be written as follows
\begin{equation}
  \label{eq:8}
2 \Pi^{(s)}_6 + \Pi^{(s)}_8 = 
2 \left(\frac{k^2}{(k\cdot u)^2}-1\right) 
\frac{\left(\Pi_6^{(s)}\right)^2}{k^2+\Pi_B^{(s)}}.
\end{equation}

The other general non-linear relations are given 
in Eqs. (\ref{eq:B13}) and (\ref{eq:B12}), which involve  quite 
lengthy expressions. We shall
present here, for simplicity, only the corresponding results
obtained in covariant gauges, at zero temperature.
In this case, the general non-linear relations simplify because in
Eq. (\ref{eq:4}), $\Pi_i = 0$ for $i\neq 4,8,12,13$. 
Furthermore, we have now $\Pi_A=\Pi_B=\Pi_C$,
since at zero temperature the only traceless-transverse
tensor available is given by the sum $T_A+T_B+T_C$ 
in (\ref{eq:9}).
Then, the condition (\ref{eq:B13}) implies that $\Pi_8=0$,
and  the non-linear constraint (\ref{eq:B12}) 
reduces to the following relation:
\begin{equation}
  \label{eq:10}
\Pi_{4} + \Pi_{12} + 2 \Pi_{13} =
\frac{3}{2 k^2}\left[\Pi_{4} \Pi_{12}-\left(\Pi_{13}\right)^2
\right]
\end{equation}
The above non-linear relations are rather interesting, since the
structure functions $\Pi_i$ are all gauge-dependent. In perturbation
theory, these functions begin at least to order $\kappa^2$.
Consequently, the linear combinations on the left-hand sides
of Eqs. (\ref{eq:8}) and (\ref{eq:10}) must
begin at order $\kappa^4$. We have verified these results
to one loop order in general axial \cite{capper:1982rc}  and 
covariant gauges, respectively, where the above linear combinations are found
to \hbox{vanish}.

\acknowledgements{We are grateful to CNPq and Fapesp, Brasil, for a grant.
J.F. would like to thank Professor J.C. Taylor for a helpful 
correspondence.} 


\appendix
\section{BRS identity}

In order to derive the BRS identity for the graviton
self-energy at finite temperature, we start from the
effective action
\begin{eqnarray}
  \label{eq:A1}
\check\Gamma & = & h_{\mu\nu}(0)\Gamma^{\mu\nu} 
 +
\int {\rm d}^4 x {\rm d}^4 y
J^{\mu\nu}(x) G_{\mu\nu}^\lambda(x-y) C_{\lambda}(y)
\nonumber \\
& + &
\frac 1 2 \int {\rm d}^4 x {\rm d}^4 yh_{\alpha\beta}(x)
\check\Gamma^{\alpha\beta,\,\mu\nu}(x-y)
h_{\mu\nu}(y)  ,
\end{eqnarray}
where $\Gamma^{\mu\nu}$ denotes the tadpole.
$J^{\mu\nu}(x)$ represents the source term for the BRS transformation
of the graviton field and $C_{\lambda}(y)$ is the vector ghost field.
The tensor $G_{\mu\nu}^\lambda(x-y)$, which is given
to lowest order
by equation (\ref{eq:3}), can be expressed 
diagrammatically in a loop expansion, as shown in references 
\cite{brandt:1998hd,capper:1982ez,delbourgo:1985wz}.

The relevant BRS invariance for the thermal graviton
self-energy can now be written as
\begin{equation}
  \label{eq:A2}
\int {\rm d}^4 x 
\frac{\delta \check\Gamma}{\delta h_{\mu\nu}(x)}
\frac{\delta \check\Gamma}{\delta J^{\mu\nu}(x)} = 0.
\end{equation}
Differentiating (\ref{eq:A2}) with respect to $C_\lambda(y)$
and setting all fields and source to zero, yields in momentum
space the following condition for the tadpole
\begin{equation}
  \label{eq:A3}
\Gamma^{\mu\nu}\left.G_{\mu\nu}^\lambda(k=0)\right|_{h=0}= 0.
\end{equation}
On the other hand, differentiating (\ref{eq:A2}) with respect to 
$C_\lambda(y)$ and $h_{\alpha\beta}(z)$, and setting the
source to zero, leads at $h=0$ to the identity
\begin{equation}
  \label{eq:A4}
\check\Gamma^{\alpha\beta,\,\mu\nu}(k) G_{\mu\nu}^\lambda(k)+
\Gamma^{\mu\nu} G_{\mu\nu}^{\lambda,\;\;\alpha\beta}(k)= 0,
\end{equation}
where $G_{\mu\nu}^{\lambda,\;\;\alpha\beta}$ denotes the
derivative of $G_{\mu\nu}^{\lambda}$ with respect to 
$h_{\alpha\beta}$. If we consider the tadpoles only for the purpose of
calculating the leading contributions to the self-energy,
Eq. (\ref{eq:A4}) can be written at
$h=0$ as follows:
\begin{equation}
  \label{eq:A5}
\left[
\check\Gamma^{\alpha\beta,\,\mu\nu}(k) G_{\mu\nu}^\lambda(k) 
\right]^{(l)} = -
\Gamma^{\mu\nu} G_{\mu\nu}^{\lambda,\;\;\alpha\beta}(k),
\end{equation}
\begin{equation}
  \label{eq:A6}
\left[
\check\Gamma^{\alpha\beta,\,\mu\nu}(k) G_{\mu\nu}^\lambda(k) 
\right]^{({ s})} = 0 .
\end{equation}
Eq. (\ref{eq:A6}) is equivalent to  (\ref{eq:7}), since 
$G_{\mu\nu}^\lambda(k)$ does not contain leading
thermal contributions. The above relations
have been verified explicitly in \cite{brandt:1998hd}, to one loop order. 
In this case, for example, Eq. (\ref{eq:A6}) becomes
(at $h=0$)
\begin{equation}
  \label{eq:A7}
\Pi^{(s)\,\alpha\beta,\,\mu\nu}(k) G_{\mu\nu}^{(0)\,\lambda}(k)+
K^{(0)\,\alpha\beta,\,\mu\nu}(k) G_{\mu\nu}^{(1)\,\lambda}(k)=0.
\end{equation}
An important consequence of (\ref{eq:A7}), which follows from the
fact that $K^{(0)}_{\alpha\beta,\,\mu\nu}(k)$ in
Eq. (\ref{eq:6}) is transverse with respect to $k_\alpha$,
is the identity
\begin{equation}
  \label{eq:A8}
k_\mu k_\alpha \Pi^{(s)\alpha\beta,\,\mu\nu}(k) = 0 .
\end{equation}
This is analogous to the identity 
$k_\mu k_\nu \Pi^{\mu\nu}(k) = 0$, which holds in QCD for the
exact gluon self-energy.

\section{Derivation of the non-linear constraints}

In this appendix, we will
derive the non-linear relations involving the elements of
$\check\Gamma^{(s)\,\mu\nu}_{\alpha\beta}$, which result
in consequence of the
vanishing of its determinant.
(In what follows, for simplicity of notation, we shall drop
the superscripts $(s)$.)
To this end, it is convenient to introduce 
a set of $10$ polarization tensors
$\epsilon^l_{\mu\nu}$ ($l=1,\cdots,10$), which constitute a
basis for the eigentensors of $\check\Gamma^{\mu\nu}_{\alpha\beta}$.
Two of these tensors describe
the physical graviton field, which has spin 2 and helicities $\pm 2$.
In terms of the transverse vectors $e^{1,2}_\mu$, which satisfy
$k^\mu e^{1,2}_\mu = u^\mu e^{1,2}_\mu =0$, we can define the physical
polarization tensors $\epsilon^{1,2}_{\mu\nu}$ as
\begin{equation}
  \label{eq:B1}
\epsilon^{1,2}_{\mu\nu}=\frac 1 2\left[
\left(e^1_\mu e^1_\nu -  e^2_\mu e^2_\nu \right)\pm
\left(e^1_\mu e^2_\nu +  e^1_\nu e^2_\mu \right)
\right],
\end{equation}
which are traceless and transverse with respect to
$k_\mu$ and $u_\mu$. These tensors are actually eigentensors of
$\check\Gamma^{\alpha\beta}_{\mu\nu}$ with the same
eigenvalue
\begin{equation}
  \label{eq:B2}
\check\Gamma_{\alpha\beta}^{\mu\nu}\; \epsilon^{1,2}_{\mu\nu} =
\left(k^2+\Pi_A\right)\;\epsilon^{1,2}_{\alpha\beta}
\end{equation}
which reflects the fact that the purely spatially transverse mode
$A$ describes a physical gravitational wave propagating in the plasma.

\noindent
Other two (unphysical) polarization tensors, defined as
\begin{equation}
  \label{eq:B3}
\epsilon^{3,4}_{\mu\nu}=
k_\mu e^{1,2}_\nu + k_\nu e^{1,2}_\mu -
\frac{k^2}{k\cdot u}\left(
u_\mu e^{1,2}_\nu + u_\nu e^{1,2}_\mu
\right),
\end{equation}
are eigentensors of $K^{(0)}_{\alpha\beta,\,\mu\nu}$ with
the same eigenvalue $k^2$.
The following two  unphysical polarization tensors can be written as
\begin{equation}
  \label{eq:B4}
\epsilon^{5,6}_{\mu\nu}=
k_\mu e^{1,2}_\nu + k_\nu e^{1,2}_\mu .
\end{equation}
These are eigentensors of $K^{(0)}_{\alpha\beta,\,\mu\nu}$ with
zero eigenvalues. In a coordinate system where $\vec k$ is directed
along the $z$-axis, they correspond to the elements of the tensor
$G^{(0)\,\,1,2}_{\mu\nu}(h=0)$ defined in Eq. (\ref{eq:3}).

The next two unphysical polarization tensors can be represented as
\begin{equation}
  \label{eq:B5}
\epsilon^{7}_{\mu\nu}=
\left(\frac{k_\mu}{k\cdot u} - u_\mu\right)k_\nu +
\left(\frac{k_\nu}{k\cdot u} - u_\nu\right)k_\mu 
\end{equation}
\begin{equation}
  \label{eq:B6}
\epsilon^{8}_{\mu\nu}= u_\mu k_\nu + u_\nu k_\mu 
\end{equation}
These are also eigentensors of $K^{(0)}_{\alpha\beta,\,\mu\nu}$
with zero eigenvalues and correspond, in the above coordinate system,
to the elements of the tensor $G^{(0)\,\,3,4}_{\mu\nu}(h=0)$.
The remaining two polarization tensors, which are given by
\begin{equation}
  \label{eq:B7}
\epsilon^{9}_{\mu\nu}= \eta_{\mu\nu} - \frac{k_\mu k_\nu}{k^2}
\end{equation}
and
\begin{eqnarray}
  \label{eq:B8}
\epsilon^{10}_{\mu\nu}&=&\epsilon^{9}_{\mu\nu} + 
\frac{3 k\cdot u}{2(k\cdot u)^2+k^2}
\nonumber \\
& \times &
\left(k_\mu u_\nu + k_\nu u_\mu - 
\frac{k^2}{k\cdot u} u_\mu u_\nu - k\cdot u \, \eta_{\mu\nu}\right),
\end{eqnarray}
are eigentensors of $K^{(0)}_{\alpha\beta,\,\mu\nu}$,
with eigenvalues $-2 k^2$ and $k^2$ , respectively.

Using Eqs. (\ref{eq:B3}) and (\ref{eq:B4}), and the properties
of the tensors $T^i_{\alpha\beta ,\,\mu\nu}$, we get the system
\begin{equation}
  \label{eq:B9}
\check\Gamma_{\alpha\beta}^{\mu\nu} \epsilon^{3,4}_{\mu\nu} =  
\left(k^2 + \Pi_B\right)  \epsilon^{3,4}_{\mu\nu} + 
2\left(1-\frac{k^2}{(k\cdot u)^2}\right)\Pi_6  \epsilon^{5,6}_{\mu\nu}
\end{equation}
\begin{equation}
  \label{eq:B10}
\check\Gamma_{\alpha\beta}^{\mu\nu} \; \epsilon^{5,6}_{\mu\nu} =   
-2\Pi_6 \; \epsilon^{3,4}_{\mu\nu}    + 
\left(4\Pi_6 + 2\Pi_8\right)\;\epsilon^{5,6}_{\mu\nu} 
\end{equation}
In order to obtain non-trivial eigentensors of 
$\check\Gamma_{\alpha\beta}^{\mu\nu}$ from the above set of
equations, we must impose the condition that the determinant
involving the corresponding eigenvalues must vanish.
This leads to a quadratic equation which 
determines a set of two eigenvalues. (Because of 
the form of Eqs. (\ref{eq:B9}) and (\ref{eq:B10}) 
the other two eigenvalues will be degenerate with those in the first
set.) The product of the eigenvalues must therefore vanish, yielding
in this way the non-linear relation given by Eq. (\ref{eq:8}).
The system (\ref{eq:B9}) and (\ref{eq:B10}) determines
altogether four eigentensors of 
$\check\Gamma_{\alpha\beta}^{\mu\nu}$, two of which have zero
eigenvalues. 

Taking into account Eq. (\ref{eq:B2}), we have thus far a total
of six eigentensors of $\check\Gamma_{\alpha\beta}^{\mu\nu}$,
so that we must still find four more eigentensors. 
These must be each linear combinations of the tensors 
$\epsilon^{l}_{\mu\nu}$ ($l=7\cdots 10$). The requirement 
of non-trivial eigentensors leads to a quartic equation for the
remaining four eigenvalues, which can be written in the form of
a vanishing determinant
\begin{equation}
  \label{eq:B11}
\left|
\begin{array}{lccr}
L_{11} - \lambda  & L_{12}  & L_{13} & L_{14} \\
L_{21}     & L_{22} - \lambda  & L_{23} & L_{24} \\ 
L_{31}     & L_{32} & L_{33} -2 k^2 - \lambda & L_{34} \\
L_{41}     & L_{42} & L_{43} & \Pi_{C} + k^2 - \lambda
\end{array}
\right| = 0 
\end{equation}
where $L_{mn}$ are linear combinations of the structure functions
$\Pi_i$ ($i=4\cdots 14$). We have, for example
\begin{eqnarray}
L_{11}  & = & 
\left.\frac{1}{r\,\left (r+1\right )}\right[
r\,\left (r-1\right )\Pi_{{4}}-
2\,\left (r-1\right )\Pi_{{5}}+
8\,r\,\Pi_{{6}}
\nonumber \\  &+&
4\,{r}^{2}\Pi_{{8}}- 
2\,\left (r-1\right )\Pi_{{9}}+
r\,\left (r-1\right )\Pi_{{12}}
\nonumber \\  &+&
2\,r\,\left (r-1\right )\Pi_{{13}}
\left. \frac {}{}   \right],
\end{eqnarray}
\begin{eqnarray}
L_{12}& = &
\left.
\frac {2\left (r-1\right )}
{\left (r+1\right )^{2}}
\right[
r\,\Pi_{{4}}+
\left (r-1\right )\Pi_{{5}}
\nonumber \\  &+&
2\,r\,\left (r-1\right )\Pi_{{6}}-
\left (r+1\right )\Pi_{{7}}+
2\,r\,\Pi_{{8}}
\nonumber \\  &+&
\left (r-1\right )\Pi_{{9}}+
r\,\left (r+1\right )\Pi_{{11}}+
r\,\Pi_{{12}}
\nonumber \\  &+&
2\,r\,\Pi_{{13}}+
r\,\left (r+1\right )\Pi_{{14}}
\left. \frac {}{}   \right],
\end{eqnarray}

\begin{eqnarray}
L_{21}&=&
\left.\frac{2}{r+1}\right[
\Pi_{{4}}+
2\,\Pi_{{5}}+
2\,\left (1+3\,r\right )\Pi_{{6}}
\nonumber \\  &+&
2\,\left (r+1\right )\Pi_{{7}}+
\left (r+3\right )\Pi_{{8}}+
2\,\Pi_{{9}}
\nonumber \\  &+&
\left (r+1\right )^{2}\Pi_{{10}}+
2\,\left (r+1\right )\Pi_{{11}}+
\Pi_{{12}}
\nonumber \\  &+&
2\,\Pi_{{13}}+
2\,\left (r+1\right )\Pi_{{14}} 
\left. \frac {}{}   \right],
\end{eqnarray}
\begin{eqnarray}
L_{22}&=&
\left.
\frac{1}{{r}^{2}}
\right[
r\,\Pi_{{4}}+\left (r-1
\right )\Pi_{{5}}+2\,r\,\left (r-1
\right )\Pi_{{6}}
\nonumber \\   &-&
\left (r+1\right )\Pi_{{7}}+
2\,r\,\Pi_{{8}}
+\left (r-1\right )\Pi_{{9}}
\nonumber \\  &+&
r\,\left (r+1\right )\Pi_{{11}}+
r\,\Pi_{{12}}+2\,
r\,\Pi_{{13}}
\nonumber \\  &+&
r\,\left (r+1
\right )\Pi_{{14}}
\left. \frac {}{}   \right],
\end{eqnarray}
where $r\equiv k^2/(k\cdot u)^2$.

As a consequence of the BRS condition given by
(\ref{eq:7}), we must have a total of four eigentensors with
zero eigenvalues. Since (\ref{eq:B9}) and (\ref{eq:B10}) have
already determined two such eigentensors, it follows that 
(\ref{eq:B11}) must have two vanishing roots. This yields
two more non-linear relations among the structure
functions, which can be written symbolically in the form
\begin{equation}
  \label{eq:B13}
k^4\left(L_{11} +  L_{22} \right) = 
k^2 L \otimes L +  L \otimes  L \otimes  L ,
\end{equation}
and
\begin{equation}
  \label{eq:B12}
k^4\left(L_{11} L_{22} - L_{12} L_{21}\right) = 
k^2 L \otimes L \otimes L  +  L \otimes  L \otimes  L \otimes L,
\end{equation}
where $L$ denotes some matrix  element of (\ref{eq:B11}).

\end{document}